\documentclass[12pt]{article} 
\usepackage{color,amsfonts,epsfig,latexsym,amscd,amsmath,theorem,mathrsfs,graphics}
\usepackage{hyperref} 

\usepackage{longtable}

\newcommand{\CC}{\mathscr{C}}

\newcommand{\evec}{{\bf e}} 
\newcommand{\fvec}{{\bf f}}

\newcommand{\R}{{\mathbb{R}}}
\newcommand{\Z}{{\mathbb{Z}}}

\newcommand{\C}{{\mathbb{C}}}

\newcommand{\LL}{{\mathsf{L}}}

\newcommand{\beq}{\begin{equation}}
\newcommand{\eeq}{\end{equation}}
\newcommand{\bea}{\begin{eqnarray}}
\newcommand{\eea}{\end{eqnarray}}
\newcommand{\ben}{\begin{eqnarray*}}
\newcommand{\een}{\end{eqnarray*}}
\newcommand{\bem}{\begin{enumerate}}
\newcommand{\eem}{\end{enumerate}}

\newcommand{\ra}{\rightarrow}

\newcommand{\cd}{\partial}

\newcommand{\wh}{\widehat}
\newcommand{\less}{\backslash}
\newcommand{\M}{{\sf M}}

\def \d{\mathrm{d}}

\newcommand{\ip}[1]{\langle #1 \rangle}
\newcommand{\ignore}[1]{}

\newcommand{\DD}{\mathscr{D}}

\newcommand{\eint}{E_{int}}

\newcommand{\ol}{\overline}

\newcommand{\eps}{\varepsilon}

\theoremstyle{plain}

{\theorembodyfont{\rmfamily}

}
\newcommand{\news}{\setcounter{equation}{0}}

\graphicspath{{./images/}} 

\begin{document}

\title{Short range intervortex forces}
\author{Martin Speight\thanks{E-mail: {\tt j.m.speight@leeds.ac.uk}}\\
School of Mathematics, University of Leeds,\\
Leeds LS2 9JT, England\\ \\
Thomas Winyard\thanks{E-mail: {\tt twinyard@ed.ac.uk}} \\
Maxwell Institute of Mathematical Sciences and School of Mathematics,\\
University of Edinburgh, Edinburgh,
EH9 3FD, United Kingdom}

%\date{}
\maketitle

\begin{abstract}
An explicit formula for the interaction energy of $n$ vortices in the abelian Higgs (or Ginzburg-Landau) model is derived, valid in the regime where all vortices are close to one another. An immediate consequence of this formula is that the interaction energy of a vortex pair with separation $d$ varies as $d^4$, not $d^2$. The formula contains $n-1$ real coefficients which are fixed by certain spectral data of the Jacobi operator of the cocentred $n$-vortex. The coefficients are computed numerically for $n=2$ and $n=3$ for couplings $0.1\leq \lambda\leq 2.5$. The resulting short range interaction potentials are compared with the results of full field theory simulations for $\lambda=0.5$ and $\lambda=2$, with excellent agreement at small to moderate vortex separation. 
\end{abstract}

\maketitle

\section{Introduction}\news

Abelian Higgs vortices are the preeminent example of topological solitons in two spatial dimensions. They have important applications in condensed matter physics and cosmology, where they arise as sections through cosmic strings. As the simplest topological soliton arising in gauge theory, they also provide an invaluable toy model for exploring soliton dynamics in the general context of high energy physics.  

To study the phenomena induced by vortices it is crucial to understand the forces between them. The abelian Higgs model on the Euclidean plane has a single dimensionless coupling constant 
$\lambda>0$ whose value controls the nature of these forces. For $\lambda<1$, vortices attract one another while for $\lambda>1$ they repel. The case of critical coupling, $\lambda=1$, is particularly well studied: here static vortices exert no net forces on one another, but there are still velocity dependent forces which can be understood within a beautiful geometric framework proposed by Manton. There is a well-developed formalism for understanding asymptotic intervortex forces at large separation, in which vortices are modelled as point sources in the linearization of the model about its vacuum \cite{spe-static}. This formalism successfully accounts for long range intervortex forces, including velocity-dependent forces at critical coupling \cite{manspe}, and has been extended to variants of the model with multiple scalar fields \cite{carbabspe,garcarbabspe,silwinbab,barsamwin} and nonlinear target space \cite{romspe,spewin-competing}.

By contrast, a quantitative understanding of {\em short} range intervortex forces has, so far, been missing. Numerical studies have indicated that, for three or more vortices, they cannot be understood as a sum of pairwise interactions \cite{cargarbab}, but a detailed analysis has been hampered by the technical challenge of computing interaction energies for close but not coincident multivortices. 

This paper aims to remedy this deficiency. It rests on two ideas. The first is to treat the $n$-vortex interaction energy $\eint$ as a function on the moduli space of $n$-vortex positions, rather than as a function of the positions directly. This distinction, while subtle, immediately yields insight: it follows, for example, that the two-vortex interaction potential varies at small separation $R$ as $R^4$, not $R^2$ as one might naively expect. The point in moduli space corresponding to $n$ coincident vortices (at the origin, say) is certainly a critical point of $\eint$, a local maximum if $\lambda>1$, a local minimum if $\lambda<1$, so the leading terms in the Taylor expansion of $\eint$ about this point are quadratic. The second idea is that we can compute the leading coefficients in this expansion by solving the eigenvalue problem for the Jacobi operator of the model associated with the coincident $n$-vortex. By rotational symmetry, this reduces to a linear ODE problem which is much more tractable than the computation of $\eint$ directly. This allows us to construct explicit formulae for the $n=2$ and $n=3$ vortex interaction potentials, valid at small separation, with coefficients whose $\lambda$ dependence we determine numerically. 

The rest of the paper is structured as follows. Section \ref{sec:model} introduces the model and defines $\eint$ precisely. In section \ref{sec:jo} we compute the Jacobi operator in a general setting and then, in section \ref{sec:sr} use the rotational equivariance of cocentred vortices in the plane to reduce it to a sequence of ordinary differential operators whose spectra are numerically amenable. This section also describes the simple shooting scheme we use to find the spectrum. Section \ref{sec:eip} explains how we can extract a small separation approximation to $\eint$ from the spectral data, and presents the numerically computed Taylor coefficients (as functions of coupling $\lambda$) for $n=2,3$. In section \ref{sec:cwft} we compute $\eint$ numerically in the full field theory via a constrained gradient descent algorithm and compare the results
with our approximate formulae for $n=2,3$ at couplings $\lambda=0.5$ and $\lambda=2$, finding excellent agreement. Section \ref{sec:cr} presents some concluding remarks.

\section{The model and its vortex interaction energy}\news\label{sec:model}

The model comprises a complex scalar field $\phi$ and a $U(1)$ gauge field $A$ on the Euclidean plane, which is conveniently identified with $\C$. To such a pair we associate the energy
\beq\label{bkw}
E(\phi,A)=\int_{\C}\left(\frac12|\d_A\phi|^2+\frac12|B|^2+\frac\lambda8(1-|\phi|^2)^2\right)
\eeq
where $B=\d A$ is the magnetic field and $\d_A\phi=\d\phi-iA\phi$ is the covariant derivative. This energy is invariant under gauge transformations,
$\phi\mapsto e^{i\chi}\phi$, $A\mapsto A+\d\chi$. To ensure finite energy, one imposes the boundary condition $|\phi|\ra 1$ as $r\ra\infty$, but the phase of $\phi$ may wind any integer number $n$ times around the unit circle as one circles the boundary at spatial infinity. Requiring $|\d_A\phi|\ra 0$ as $r\ra\infty$, a standard invocation of Stokes's theorem implies that the total magnetic flux of a winding $n$ configuration is
$\int_\C B=2\pi n$. Without loss of generality, we may assume that $n\geq 0$. The Higgs field of a winding $n$ configuration (generically) vanishes at $n$ points in $\C$ counted with multiplicity, which we interpret as vortex positions (or antivortex positions if their multiplicity is negative). 

For each $n\in\Z^+$ there is a rotationally symmetric static solution of the model (that is, critical point of $E$) consisting of $n$ vortices colocated at the origin, which we will call the $n$-vortex. In suitable gauge this takes the form
\beq\label{ansatz}
\phi=f(r)e^{in\theta},\qquad A=a(r)\d\theta,
\eeq
where $f,a:(0,\infty)\ra\R$ satisfy
\bea
-f''-\frac{f'}{r}+(a-n)^2\frac{f}{r^2}+\frac\lambda2(f^2-1)f&=&0\\
-a''+\frac{a'}{r}+(a-n)f^2&=&0,
\eea
subject to the boundary conditions $f(0)=a(0)=0$, $f(\infty)=1$, $a(\infty)=n$. 
For $\lambda<1$ this solution is stable (a local minimum of $E$) while for
$\lambda>1$ it is unstable (a saddle point). If $\lambda=1$ it is one point in a $2n$ dimensional family of static solutions all of equal energy, but for $\lambda\neq 1$, this solution is thought to be unique up to gauge and translation. 

Given a choice of $n$ points $z_1,\ldots,z_n$ in $\C$, possibly with repeats, there is no static solution with vortices located at the points $z_i$, unless they are coincident (or $\lambda=1$): there are forces between static vortices, encapsulated by their interaction energy. This assigns, to the collection of marked points 
$z_1,\ldots,z_n$,
\beq
\eint(z_1,\ldots,z_n):=\inf E(\phi,A) -n E_1
\eeq
where the infimum is over all smooth fields vanishing at exactly the points $z_i$ (with the correct multiplicity), and having winding $n$ at infinity, and $E_1$ is the energy of a single vortex. This infimum is attained only in the coincident case, by (a translate of) the $n$-vortex \eqref{ansatz}. If $\lambda<1$ ($\lambda>1$) it is strictly
negative (positive). 

It is clear that the order in which we label the marked points $z_i$ is irrelevant, so $\eint$ is actually a function on $\C^n/S_n$, the quotient of $\C^n$ by the symmetric group, that is the symmetric $n$-fold product of $\C$. Although this space is a smooth manifold diffeomorphic to $\C^n$ itself, it is helpful to give it a different name, to emphasize that this is {\em not} the space of ordered vortex positions $(z_1,z_2,\ldots,z_n)$. We will denote it $\M_n$, and call it the $n$ vortex configuration space. To see that $\M_n\equiv\C^n$, we identify the permutation orbit of
$(z_1,z_2,\ldots,z_n)$ with the unique monic polynomial whose roots are  
$z_1,z_2,\ldots,z_n$:
\beq
p(z)=(z-z_1)(z-z_2)\cdots(z-z_n)=:z^n +a_1z^{n-1}+a_2z^{n-2}+\cdots+a_{n-1}z+a_n.
\eeq
Hence, the interaction energy is actually a function of $a\in\C^n$ (the coefficients of this polynomial).  More abstractly, we think of $a_i$ as global complex coordinates on the space $\M_n$.

By translation symmetry, we may restrict to the set of $n$-vortex configurations whose centre of mass is at $z=0$, that is, satisfying
\beq
z_1+z_2+\cdots+z_n=-a_1=0,
\eeq
so $E_{int}:\C^{n-1}\ra\R$. Again, it is conceptually helpful to denote this submanifold of $\M_n$ by $\M_n^0$, and call it the centred $n$ vortex configuration space. We assume that $E_{int}(a_2,a_3,\ldots,a_n)$ is smooth (or at least twice differentiable).
The radially symmetric $n$-vortex corresponds to
$p(z)=z^n$, that is, $a=0$, and is a critical point of $E_{int}$ --
a local maximum if $\lambda>1$ and a local minimum if $\lambda<1$. Hence, provided all the vortices are close to $0$, $\eint(a)$ should be well approximated by its Taylor expansion about $a=0$ to quadratic order in $a$, that is,
\beq\label{nava}
\eint(a)=\eint(0)+\frac12\sum_{i,j=2}^n(M_{ij}a_ia_j+
H_{ij}\ol{a_i}a_j+\ol{M_{ij}}\ol{a_i}\ol{a_j})+\cdots
\eeq
where $M,H$ are complex $(n-1)\times (n-1)$ matrices and $H$ is hermitian.
Note the expansion has no linear terms since $a=0$ is a critical point of $\eint$. Note also that $\eint(0)=E_n-nE_1$, where $E_n$ is the energy of the rotationally symmetric $n$-vortex. 

The interaction energy is invariant under simultaneous rotation of all the
 vortex positions. That is, for all $w\in U(1)$, the map $z_i\mapsto wz_i$
  is a symmetry of $\eint$ (preserving the centring condition $a_1=0$).
   The action of this map on the polynomial $p(z)$ is $p(z)\mapsto
w^np(z/w)$, so maps the coefficients $a_i\mapsto w^i a_i$. Hence
$\eint(a)$ is invariant under the $U(1)$ action
\beq
(a_2,\ldots, a_n)\mapsto w\cdot a:=(w^2a_2,\ldots,w^na_n).
\eeq
Since $\eint(w\cdot a)=\eint$ for all $a\in\C^{n-1}$ and $w\in U(1)$, we see from \eqref{nava} that
\beq
\sum_{i,j}(w^{i+j}M_{ij}a_ia_j+w^{j-i}H_{ij}\ol{a_i}a_j
+w^{-(i+j)}\ol{M_{ij}}\ol{a_i}\ol{a_j})
=\sum_{i,j}(M_{ij}a_ia_j+
H_{ij}\ol{a_i}a_j+\ol{M_{ij}}\ol{a_i}\ol{a_j})
\eeq
for all $a$ and $w$, so $M=0$ and $H$ is diagonal (and hence real). That is, there exist real numbers $c_2,c_3,\ldots,c_n$ such that
\beq\label{brkawi}
\eint(a)=(E_n-nE_1)+\frac12\sum_{k=2}^n c_k|a_k|^2+O(|a|^3).
\eeq
The real coefficients $c_k$ depend on $\lambda$, are all positive for $\lambda<1$, all negative for $\lambda>1$, and all vanish at $\lambda=1$. To complete our short-range approximation to $\eint$ we must compute them. To do so, we will consider the second variation of $E$ about the $n$-vortex \eqref{ansatz}.

\section{The Jacobi operator}\news\label{sec:jo}

Assume we have a static solution $(\phi,A)$ of this model, i.e.\ a critical point of $E$. We wish to understand the second variation of $E$ about
$(\phi,A)$, which is encoded in the spectral properties of its associated Jacobi operator $J$. The spectrum of $J$ has been heavily studied before, from the original work of Goodband and Hindmarsh \cite{goohin} to more recent detailed studies by Alonso-Izquierdo and collaborators \cite{alomig,alomigque}. We will require not just the low-lying eigenvalues of $J$, but also their associated eigenmodes, which cannot be read off from previous work. We have no choice, therefore, but to solve the eigenvalue problem for $J$ afresh and, this being the case, we take the opportunity to give a more geometric derivation of $J$ and its symmetry reduction than has appeared previously. 

For this purpose, it is helpful to think of $\phi$ as a section of a Hermitian line bundle $\LL$, with inner product
$h(\phi,\psi)=(\ol\phi\psi+\phi\ol\psi)/2$,  over a Riemannian $2$-manifold $\Sigma$, and $A$ as a connexion on $\LL$.  We will revert to the choice of direct interest, $\Sigma=\R^2$, in section \ref{sec:sr}.

Consider a two-parameter variation $(\phi_{s,t},A_{s,t})$ of $(\phi,A)=(\phi_{0,0},A_{0,0})$ and define the infinitesimal perturbations it generates
\beq
\eps=\frac{d\:}{ds}\bigg|_{s=0}\phi_{s,0},\quad
\wh\eps=\frac{d\:}{dt}\bigg|_{t=0}\phi_{0,t},\quad
\alpha=\frac{d\:}{ds}\bigg|_{s=0}A_{s,0},\quad
\wh\alpha=\frac{d\:}{dt}\bigg|_{t=0}A_{0,t}.
\eeq
Note that $\eps,\wh\eps$ are, like $\phi$, sections of $\LL$, while $\alpha,\wh\alpha$ are (globally defined) one-forms on $\Sigma$. Then
\bea
\frac{\cd^2 E(\phi_{s,t},A_{s,t})}{\cd s\cd t}\big|_{(s,t)=(0,0)}&=& {\rm Hess}((\wh\eps,\wh\alpha),(\eps,\alpha))\nonumber\\
&=&\ip{\wh\eps,\Delta_A\eps+\frac\lambda2(h(\phi,\phi)-1)\eps+\lambda h(\phi,\eps)\phi}_{L^2}\nonumber \\
&&+
\ip{\wh{\eps},i*(\alpha\wedge*\d_A\phi+\d_A(*\alpha\phi))}_{L^2}\nonumber \\
&&
+\ip{\wh\alpha,h(\eps,i\d_A\phi)+h(\phi,i\d_A\eps)}_{L^2}\nonumber \\
&&
+\ip{\wh{\alpha},\delta\d\alpha+h(\phi,\phi)\alpha}_{L^2}.
\eea
In this formula, $\ip{\cdot,\cdot}_{L^2}$ denotes $L^2$ inner product,
$\Delta_A$ is the gauge covariant Laplacian, $\Delta_A=-*\d_A*\d_A$, and $\delta$ is the coderivative adjoint to $\d$. 

From this symmetric bilinear form, we extract the {\em Jacobi operator} for the solution $(\phi,A)$,
\beq
J\left[\begin{array}{c}\eps\\\alpha\end{array}\right]=
\left[\begin{array}{c}\Delta_A\eps+\frac\lambda2(|\phi|^2-1)\eps+\lambda h(\phi,\eps)\phi+i*(\alpha\wedge *\d_A\phi+\d_A(*\alpha\phi))\\
\delta\d\alpha+|\phi|^2\alpha+h(\eps,i\d_A\phi)+h(\phi,i\d_A\eps)\end{array}\right],
\eeq
defined by the requirement that 
\beq
{\rm Hess}((\wh\eps,\wh\alpha),J(\eps,\alpha))=\ip{(\wh\eps,\wh\alpha),J(\eps,\alpha)}_{L^2}.
\eeq
This is a formally self-adjoint operator on $\Gamma(L)\oplus\Omega^1(\Sigma)$, with respect to its natural $L^2$ inner product, that is,
\beq
\ip{(\wh\eps,\wh\alpha),J(\eps,\alpha)}_{L^2}\equiv
\ip{(\eps,\alpha),J(\wh\eps,\wh\alpha)}_{L^2}.
\eeq
This follows immediately from  the symmetry of ${\rm Hess}$, but can also be verified by explicit calculation.
The spectrum of $J$ informs us about the stability of the critical point $(\phi,A)$: if the spectrum is non-negative, the solution is linearly stable. Eigensections with negative eigenvalue are perturbations which decrease $E$ to second order and hence constitute directions of instability. 

Any $(\eps,\alpha)$ tangent to a deformation which does not change $E$ should be in the kernel of $J$. For example, $E$ is gauge invariant, so all infinitesimal gauge transformations
\beq
(\eps,\alpha)=(i\phi\chi,\d\chi)
\eeq
where $\chi:\Sigma\ra\R$ is an arbitrary smooth function, are in $\ker J$.  

Since all infinitesimal gauge transformations are in $\ker J$, this kernel is infinite dimensional. Let $(\eps,\alpha)$ be any eigensection of $J$ with eigenvalue $\Lambda\neq 0$. Then, since $J$ is self adjoint, for any $(\wh\eps,\wh\alpha)\in\ker J$,
\beq
\ip{(\wh\eps,\wh\alpha),(\eps,\alpha)}_{L^2}=\frac{1}{\Lambda}\ip{(\wh\eps,\wh\alpha),J(\eps,\alpha)}_{L^2}
=\frac{1}{\Lambda}\ip{J(\wh\eps,\wh\alpha),(\eps,\alpha)}_{L^2}=0.
\eeq
Hence, every such eigensection is $L^2$ orthogonal to all infinitesimal gauge transformations. We may therefore insist that $(\eps,\alpha)$ is $L^2$ orthogonal to
the subspace
\beq
G_\infty:=\{(i\phi\chi,\d\chi)\: :\: \chi\in C^\infty(\Sigma,\R)\}.
\eeq
Then $(\eps,\alpha)$ must satisfy the PDE
\beq\label{cancol}
\delta\alpha+h(\eps,i\phi)=0.
\eeq

In the case of interest, $\Sigma=\R^2$, translation is also a symmetry so, for example
\beq\label{transmode}
(\eps,\alpha)=(\cd_x\phi,\cd_xA)
\eeq
is in $\ker J$. 
Note that this translational zero mode does not (necessarily) satisfy the gauge orthogonality condition \eqref{cancol}.

\section{Symmetry reduction}\news\label{sec:sr}

From now on, assume that $\Sigma=\R^2\equiv\C$ and that $(\phi,A)$ is the cocentred $n$-vortex solution \eqref{ansatz}. In this section we
rederive the decomposition of $J$ into a sequence of ordinary differential operators observed in \cite{alomig,alomigque}, clarifying how this results directly from the equivariance of the $n$-vortex with respect to rotations and reflexions.

Given $w\in U(1)$ denote by the same symbol the rotation map $\C\ra\C$, $z\mapsto wz$. Then the $n$-vortex is
invariant under the circle action
\beq
\phi\mapsto w^{-n}\phi\circ w,\qquad
A\mapsto w^*A.
\eeq
(The symbol $w^*$ in the above formula denotes the pullback of $A$ by the map $w$. We will always denote complex conjugation by an overbar.) This circle action is a symmetry of the functional $E$, so $J$ must preserve the invariant subspaces of its action on $\Gamma(L)\oplus\Omega^1(\Sigma)$. These are labelled by $k\in\{0,1,2,\ldots\}$,
\bea
\CC_k&=&\{\eps_+(r)e^{i(n+k)\theta}+\eps_-(r)e^{i(n-k)\theta}\: :\: \eps_{\pm}:(0,\infty)\ra\C\}\oplus \\
 &&\{
(\alpha_1(r)\cos k\theta+\alpha_2(r)\sin k\theta)\d r+
(\alpha_3(r)\cos k\theta +\alpha_4(r)\sin k\theta)r\d\theta\: :\: 
\alpha_a:(0,\infty)\ra\R\}.\nonumber
\eea

The vortex is also invariant under the parity operation $\Pi:(\phi,A)\mapsto (c\circ\phi\circ c,-c^*A)$, where $c:\C\ra\C$ is complex conjugation. Each of the subspaces $\CC_k$ decomposes further into a pair of invariant subspaces, preserved by $\Pi$
\bea
\CC_k^+&=&\{\eps_1(r)e^{i(n+k)\theta}+\eps_3(r)e^{i(n-k)\theta},\alpha_2(r)\sin k\theta\, \d r+
\alpha_3(r)\cos k\theta\, r\d\theta\}\\
\CC_k^-&=&\{i\eps_2(r)e^{i(n+k)\theta}+i\eps_4(r)e^{i(n-k)\theta},\alpha_1(r)\cos k\theta\, \d r+
\alpha_4(r)\sin k\theta\, r\d\theta\}
\eea
where $\eps_a:(0,\infty)\ra\R$. So $\CC_k^+$ is the subspace on which $\eps_\pm$ are real and $\alpha_1=\alpha_4=0$, while $\CC_k^-$ is the subspace on which $\eps_{\pm}$ are imaginary and $\alpha_2=\alpha_3=0$.

$J$ preserves this splitting, that is  $J:\CC_k^+\ra\CC_k^+$ and $J:\CC_k^-\ra\CC_k^-$. In particular, its action on $\CC_k^+$ is
\beq\label{cahu}
J_k^+\left[\begin{array}{c}\eps_1\\ \eps_3\\ \alpha_2 \\ \alpha_3\end{array}\right]=
\left[\begin{array}{c}
\DD_{n+k}\eps_1+\frac\lambda2(f^2-1)\eps_1+\frac\lambda2f^2(\eps_1+\eps_3)+\alpha_2 f'-\alpha_3(n+\frac{k}2-a)\frac{f}{r}+\frac{f(r\alpha_2)'}{2r}\\
\DD_{n-k}\eps_3+\frac\lambda2(f^2-1)\eps_3+\frac\lambda2f^2(\eps_1+\eps_3)-\alpha_2 f'-\alpha_3(n-\frac{k}2-a)\frac{f}{r}-\frac{f(r\alpha_2)'}{2r}\\
-\frac{k}{r}\left(-\frac{k}{r}\alpha_2+\frac{(r\alpha_3)'}{r}\right)+f^2\alpha_2+f'(\eps_1-\eps_3)-f(\eps_1'-\eps_3')\\
-\left(-\frac{k}{r}\alpha_2+\frac{(r\alpha_3)'}{r}\right)'+f^2\alpha_3-2(n-a)\frac{f}{r}(\eps_1+\eps_3)-\frac{k}{r}f(\eps_1-\eps_3)
\end{array}\right]
\eeq
where, for any integer $q$,
\beq
\DD_q\xi:=-\xi''-\frac{\xi'}{r}+\frac{(q-a(r))^2}{r^2}\xi.
\eeq
A section in $\CC_k^+$ is gauge orthogonal (satisfies \eqref{cancol}) if and only if
\beq\label{carhun}
\alpha_2'=-\frac{\alpha_2}{r}+\frac{k\alpha_3}{r}+f(\eps_1-\eps_3).
\eeq

For each $k\geq 1$, the linear map
$
L:\CC_k^+\ra\CC_k^-$,
\bea
&&L:(\eps_1e^{i(n+k)\theta}+\eps_3e^{i(n-k)\theta},\alpha_2\sin k\theta \d r+
\alpha_3\cos k\theta r \d\theta)\mapsto\nonumber \\
&&\qquad(i\eps_1e^{i(n+k)\theta}-i\eps_3e^{i(n-k)\theta},\alpha_2\cos k\theta \d r-
\alpha_3\sin k\theta r \d\theta)\label{jg}
\eea
commutes with $J$. Also, if $(\eps,\alpha)\in \CC_k^+$ satisfies the gauge orthogonality condition \eqref{carhun}, then $L(\eps,\alpha)\in\CC_k^-$ satisfies the gauge orthogonality condition on $\CC_k^-$,
\beq\label{gperp-}
\alpha_1'=-\frac{\alpha_1}{r}-k\frac{\alpha_4}{r}+f(\eps_2+\eps_4).
\eeq
It follows that if $v=(\eps,\alpha)\in\CC_k^+$ is an eigensection of $J$ with eigenvalue $\Lambda$, so is $Lv\in\CC_{k}^-$. So eigensections come in degenerate pairs, and we may restrict attention to $\CC_k^+$. The subspace $\CC_0$ is exceptional since $\CC_0=\CC_0^+\oplus G_\infty$, that is, the $-1$ eigenspace of $\Pi$ consists of infinitesimal gauge transformations. Hence, eigensections with $k=0$ do not come in pairs, but it is still true that one need only consider the subspace $\CC_0^+$. For our purposes, we will need $\CC_k^+$ for $k=2,3,\ldots,n$ only.

To construct an eigensection of $J$ in the class $\CC_k^+$, we must solve the ODE system
\beq\label{odes}
J_k^+\left[\begin{array}{c}\eps_1\\ \eps_3\\ \alpha_2 \\ \alpha_3\end{array}\right]=
\Lambda\left[\begin{array}{c}\eps_1\\ \eps_3\\ \alpha_2 \\ \alpha_3\end{array}\right]
\eeq
coupled to the gauge orthogonality condition \eqref{carhun}, and for this we must determine the correct boundary conditions for $(\eps_1,\eps_3,\alpha_2,\alpha_4)$.
All the components $\eps_1,\eps_3,\alpha_2,\alpha_3$ should approach $0$ exponentially fast as $r\ra\infty$. Their boundary behaviour at $0$ is determined by demanding that the section $\eps$ and the one-form $\alpha$ should be smooth at the origin. So
\beq
\eps_1\sim r^{n+k},\qquad \eps_3\sim r^{|n-k|}.
\eeq
The boundary conditions for $\alpha_2,\alpha_3$ are more subtle. Noting that $\d r= \cos\theta \d x+ \sin\theta \d y$ and
$r\d\theta=-\sin\theta \d x+\cos\theta \d y$, we find
\beq
\alpha=[\alpha_2(r)\sin k\theta\cos\theta-\alpha_3(r)\cos k\theta\sin\theta]\d x
+[\alpha_2(r)\sin k\theta\sin\theta+\alpha_3(r)\cos k\theta\cos\theta]\d y
\eeq
so $\alpha$ is smooth at the origin if and only if the functions $A_1,A_2:\R^2\less\{(0,0)\}\ra\R$
\bea
A_1&:=&\alpha_2(r)\sin k\theta\cos\theta-\alpha_3(r)\cos k\theta\sin\theta\nonumber\\
&=&\frac12(\alpha_2(r)-\alpha_3(r))\sin(k+1)\theta+\frac12(\alpha_2(r)+\alpha_3(r))\sin(k-1)\theta\\
A_2&:=&\alpha_2(r)\sin k\theta\sin\theta+\alpha_3(r)\cos k\theta\cos\theta
\nonumber\\
&=&-\frac12(\alpha_2(r)-\alpha_3(r))\cos(k+1)\theta+\frac12(\alpha_2(r)+
\alpha_3(r))\cos(k-1)\theta
\eea
extend smoothly to $(x,y)=(0,0)$. This requires that
\beq
\alpha_2-\alpha_3\sim r^{k+1},\qquad 
\alpha_2+\alpha_3\sim r^{|k-1|}.
\eeq
Hence, at small $r$,
\bea
\eps_1(r)&=&e_1r^{n+k}+\cdots\nonumber \\
\eps_3(r)&=&e_2r^{|n-k|}+\cdots\nonumber \\
\alpha_2(r)&=&e_3r^{|k-1|}+e_4r^{k+1}+\cdots\nonumber \\
\label{asgikias}
\alpha_3(r)&=&e_3r^{|k-1|}-e_4r^{k+1}+\cdots
\eea
for some unknown constants $e_1,e_2,e_3,e_4\in\R$.

Our computational scheme may be described as follows. We reinterpret the 1st, 2nd and 4th equations in \eqref{odes}, together with \eqref{carhun}, 
\bea
\DD_{n+k}\eps_1+\frac\lambda2(f^2-1)\eps_1+\frac\lambda2f^2(\eps_1+\eps_3)+\alpha_2 f'-\alpha_3(n+\frac{k}2-a)\frac{f}{r}+\frac{f(r\alpha_2)'}{2r}-\Lambda\eps_1&=&0\nonumber \\
\DD_{n-k}\eps_3+\frac\lambda2(f^2-1)\eps_3+\frac\lambda2f^2(\eps_1+\eps_3)-\alpha_2 f'-\alpha_3(n-\frac{k}2-a)\frac{f}{r}-\frac{f(r\alpha_2)'}{2r}
-\Lambda\eps_3&=&0 \nonumber \\
-\left(-\frac{k}{r}\alpha_2+\frac{(r\alpha_3)'}{r}\right)'+f^2\alpha_3-2(n-a)\frac{f}{r}(\eps_1+\eps_3)-\frac{k}{r}f(\eps_1-\eps_3)-\Lambda\alpha_3 &=&0\nonumber \\
\alpha_2'+\frac{\alpha_2}{r}-\frac{k\alpha_3}{r}-f(\eps_1-\eps_3)&=&0 \nonumber \\ &&
\label{agka}
\eea
as a first order flow for the collected fields
\beq
\phi = (\eps_1,\eps_3,\alpha_3,\eps_1',\eps_3',\alpha_3',\alpha_2)
\eeq
in $\R^7$, which we 
solve using a shooting method. We choose $r_0 \ll 1$ and $r_2 \gg 1$, then shoot forwards from $r=r_0$ and backwards from $r=r_2$, matching at $r_1 = (r_0 + r_2 )/2$. Let us denote by $S_0:\R^4\ra\R^7$ the linear map 
\beq
S_0:(e_1,e_2,e_3,e_4)\mapsto\phi_0(r_1)
\eeq
where $\phi_0:[r_0,r_1]\ra\R^7$ is the solution of \eqref{agka} with initial data $\phi_0(r_0)$ as determined by the asymptotic expressions in \eqref{asgikias} evaluated at $r=r_0$. So $S_0$ maps the left shooting data to the value of the solution $\phi$ at the matching point $r_1$. We similarly define the right shooting map $S_2:\R^3\ra\R^7$ by
\beq
S_2:(b_1,b_2,b_3)\mapsto\phi_2(r_1)
\eeq
where $\phi_2:[r_1,r_2]\ra\R^7$ is the solution of \eqref{agka} with final data
\beq
\phi_2(r_2)=(0,0,0,-r_2b_1e^{-r_2},-r_2b_2e^{-r_2},-r_2b_3e^{-r_2},0).
\eeq
So we approximate the decaying boundary condition as $r\ra\infty$ by imposing that $\eps_i,\alpha_i$ vanish at some large fixed $r_2$, with exponentially small derivatives. Again $S_2$ is linear by linearity of \eqref{agka}. In practice, we construct $S_1,S_2$ by solving \eqref{agka} numerically using a 4th order Runge-Kutta method. 

Now $\Lambda$ is an eigenvalue of $J_k^+$ if and only if there exist nonzero $e=(e_1,e_2,e_3,e_4)\in\R^4$ and $b=(b_1,b_2,b_3)\in\R^3$ such that $S_0(e)=S_2(b)$; the corresponding eigenfunction is then the solution of \eqref{agka} with shooting data $e$ at $r_1$ and $b$ at $r_2$. To determine whether such a pair $(e,b)\in\R^4\oplus\R^3$ exists we construct the $7\times 7$ matrix $Q(\Lambda)$
\beq
Q(\Lambda)=\left(\begin{array}{ccccccc}
\uparrow&\uparrow&\uparrow&\uparrow&\uparrow&\uparrow&\uparrow\\
S_0(\evec_1)&S_0(\evec_2)&S_0(\evec_3)&S_0(\evec_4)&
-S_2(\fvec_1)&-S_2(\fvec_1)&-S_2(\fvec_3)\\
\downarrow&\downarrow&\downarrow&\downarrow&\downarrow&\downarrow&\downarrow\\
\end{array}\right)
\eeq
where $\{\evec_i\}$, $\{\fvec_i\}$ are the usual bases for $\R^4$ and $\R^3$ respectively. A smooth solution of the shooting problem with parameter $\Lambda$ exists if and only if $Q(\Lambda)$ has nontrivial kernel, that is, if and only if $\det Q(\Lambda)=0$. So we compute $Q(\Lambda)$ as a function of $\Lambda$ and then solve $\det Q(\Lambda)=0$ using the bisection method. Having identified $\Lambda$, we construct $(e,b)\in\ker Q(\Lambda)\subset\R^7$. The corresponding eigenfunction is then the solution with shooting data $e$ at $r_0$ and $b$ at $r_2$. 

In principle, this method can be used to find any and all eigenvalues of $J$, together with their corresponding eigensections. We will only need those eigenmodes whose eigenvalues $\Lambda(\lambda)$ pass through $0$ at $\lambda=1$. Let us denote the space of such eigenmodes $V(\lambda)$. It has dimension $2n$ and is spanned by one vector in $\CC_k^+$ for each of 
$k=1,2,\ldots,n$, together with their images under $L$ (see \eqref{jg}).
The numerical results of finding $\Lambda(\lambda)$ using the method above are shown in figure \ref{fig:coefs} for $n = 2,3$. As a consistency check, we have also computed numerically the lowest eigenvalue in the space $\CC_1^+$. This coincides with the (gauge orthogonal component of) the overall translation mode in the $x$-direction \eqref{transmode}, so must be $\Lambda_1\equiv 0$ for all $\lambda$. We find numerically that $|\Lambda_1(\lambda)|<10^{-5}$ for $ 1/4 \leq\lambda\leq 2$, which gives an indication of the expected accuracy of our results. 

\section{Extracting the $n$-vortex interaction potential}\news\label{sec:eip}

Our short range approximation to $\eint:\M_n\ra\R$ amounts to replacing it by its
 Hessian at $0$, a symmetric bilinear form on the tangent space to $\M_n$ at $0$. At 
 $\lambda=1$, we may identify this tangent space with $\ker J$ or, more precisely, 
 the $2n$-dimensional subspace of $\ker J$ orthogonal to $G_\infty$. This, in 
 the notation of section \ref{sec:sr}, is $V(1)$. For $\lambda\neq 1$, this
  identification persists: we may identify the tangent space to $\M_n$ at $0$ 
  with $V(\lambda)$, the $2n$-dimensional subspace spanned by the eigenmodes of 
  $J$ whose eigenvalues pass through $0$ at $\lambda=1$. Having made this
   identification, it is natural to posit that the Hessian of $\eint $ at $0$ 
   coincides with the restriction to $V(\lambda)$ of the Hessian of the Ginzburg-Landau energy
    functional \eqref{bkw} at the $n$-vortex. We will test this
     supposition numerically in section \ref{sec:cwft}. This allows us to extract 
     the coefficients $c_k$ in our short range formula \eqref{brkawi} for $\eint$ 
     from spectral data for $J$.

To be explicit, choose $k\in\{2,3,\ldots,n\}$ and consider the eigenmode $v=(\eps,\alpha)\in V(\lambda)\cap \CC_k^+$, normalized so that $\|v\|_{L^2}=1$. Denote by $\Lambda_k$ its eigenvalue and by $b_k$ its associated left shooting coefficient $e_2$ (see \eqref{asgikias}). Consider now the curve of configurations
\beq
(\phi_t,A_t)=(\phi,A)+tv=(\phi+t\eps,A+t\alpha).
\eeq
This is a smooth curve passing through the symmetric $n$-vortex
\beq
(\phi,A)=(f(r)e^{in\theta},a(r)\d\theta),
\eeq
so
\beq
\frac{d\:}{dt}\bigg|_{t=0}E(\phi_t,A_t)=0,
\eeq
and, by the definition of the Jacobi operator,
\beq\label{skhe}
\frac{d^2\:}{dt^2}\bigg|_{t=0}E(\phi_t,A_t)=\ip{v,Jv}_{L^2}=\Lambda_k\|v\|_{L^2}^2=\Lambda_k.
\eeq

We wish to identify the curve $(\phi_t,A_t)$ with a curve in the centred $n$-vortex configuration space $a(t)\in\C^{n-1}$. To do so, we must identify the monic polynomial whose roots coincide with the zeros of $\phi_t(z)$.  For small $t$, these roots will be close to $0$, where the small $z$ expansions
\beq
\phi(z)=f_0z^n+\cdots,\qquad
\eps(z)=b_kz^{n-k}+\cdots
\eeq
are valid, where
\beq
f(r)=f_0r^n+\cdots,\qquad \eps_3(r)=b_kr^{n-k}+\cdots.
\eeq
So
\beq
\phi_t(z)=f_0(z^n+\frac{b_k}{f_0}tz^{n-k})+\cdots,
\eeq
that is, for small $t$, the curve $\phi_t$ corresponds to the curve of polynomials with
\beq
a_j(t)=\left\{\begin{array}{cc} 0, & j\neq k, \\
{b_k}t/f_0, & j=k.\end{array}\right.
\eeq
The second derivative of $\eint$ along $a(t)$ is
\beq
\frac{d^2\:}{dt^2}\bigg|_{t=0}\eint(a(t))=c_k\frac{b_k^2}{f_0^2}.
\eeq
Matching with \eqref{skhe}, we find that
\beq
c_k=(f_0/b_k)^2\Lambda_k.
\eeq

\begin{figure}
\begin{center}
\includegraphics[width=\linewidth]{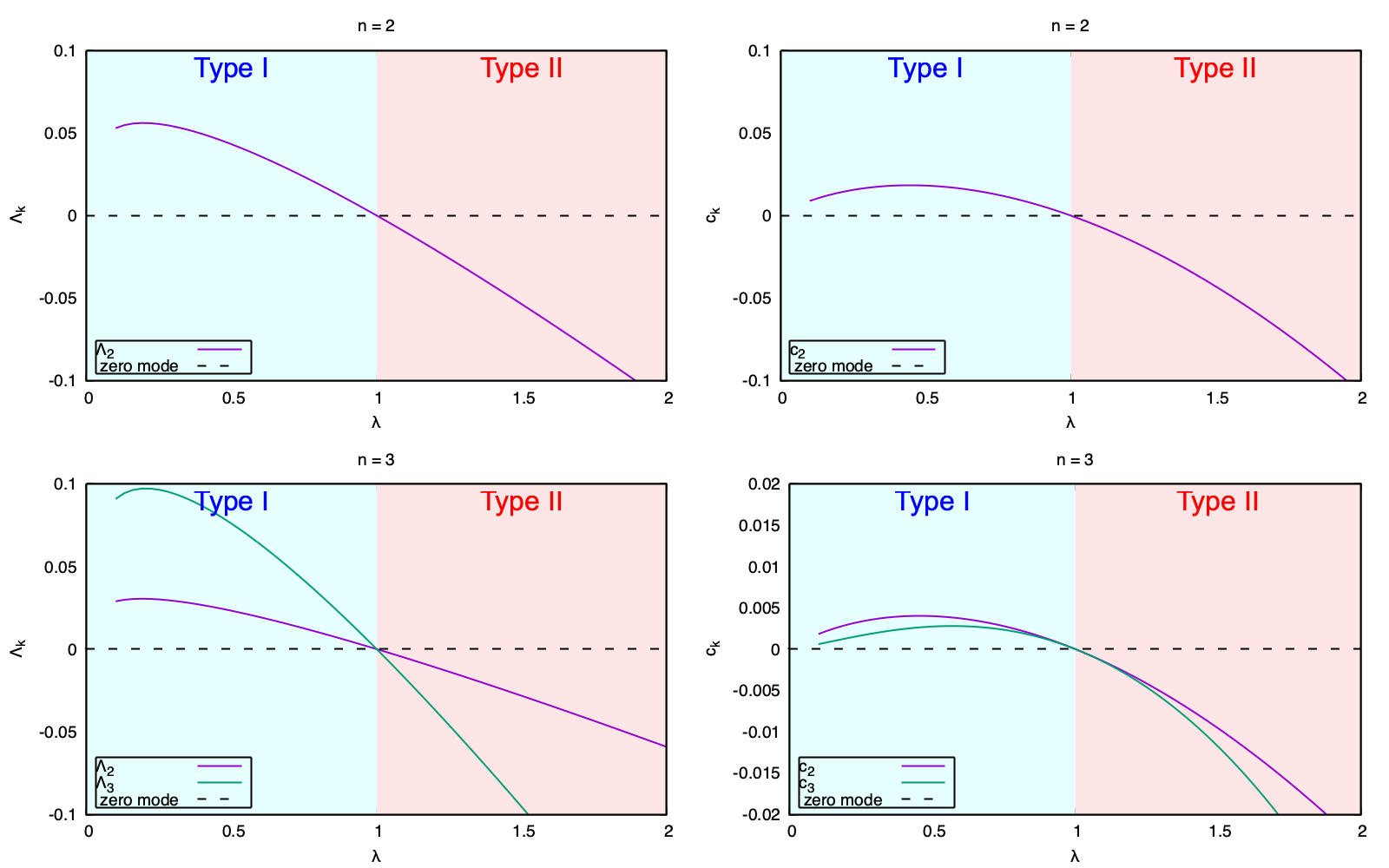}
\caption{Plots of the eigenvalues of $J$ (left column) and the coefficients of the short range approximation for the $n$-vortex interaction energy (right column) for $n=2,3$.}
\label{fig:coefs}
\end{center}
\end{figure}

To summarize, our short range approximation to the $n$-vortex interaction energy is
\beq
\eint^{(0)}(a)=E_n-nE_1+\frac{f_0^2}{2}\sum_{k=2}^n\frac{\Lambda_k}{b_k^2}|a_k|^2,
\label{eq:short}
\eeq
where $f_0$ is the leading coefficient of the expansion of the vortex profile function $f(r)$ about $r=0$, $\Lambda_k$ is the eigenvalue of the eigenmode in $\CC_k^+$ which passes through $0$ at $\lambda=1$, and $b_k$ is the leading coefficient of the expansion of $\eps_3(r)$ about $r=0$ for this ($L^2$ normalized) eigenmode. Plots of the coefficients $c_k(\lambda)$ for $n=2,3$ are presented in figure \ref{fig:coefs}.

\section{Comparison with field theory}\news\label{sec:cwft}

We now compare the predictions for the short range interaction energy with a direct computation of $\eint$ in numerical field theory. Our algorithm is explained in detail in \cite{spewin-competing}: we minimize $E$ over all fields on a large rectangle, with $\phi$ having winding $n$ on the boundary, subject to the constraint that $\phi=0$ at a collection of $n$ prescribed points in the rectangle. In practice, this is achieved by solving Newton flow for a lattice approximant to $E$ with an arresting criterion which sets the velocity of the fields to $0$ if the flow starts to move opposite to the direction of the gradient\cite{spewin}. The Higgs field at the prescribed points is simply fixed to $0$. The results presented below were all obtained using a lattice of size $N_1\times N_2 = 1001 \times 1001$ and mostly with equal lattice spacings $h_1=h_2=0.05$ (vortex pairs and collinear vortex triples). To compute the interaction energy of an equilateral {\em triangle} of vortices, it is more convenient to use a rectangular but not square lattice. Choosing $h_2=\sqrt{3}h_1$, our lattice can accommodate vortices positioned at the sites $(0,0)$, $(2mh_1,0)$, $(mh_1,2mh_2)$ for any positive integer $m$, and these form the vertices of an equilateral triangle. The interaction energies of vortex triangles were computed on such a lattice with $h_1=0.05$.

The derivatives were approximated using a 4th order central finite difference scheme. The ``time evolution" of the Newton flow was implemented via the Euler method with timestep $\delta t=h_1h_2$ and, as in \cite{spewin-competing}, a force arresting criterion was used. Since full field theory simulations are computationally costly, we construct $\eint$ only for two representative choices of coupling, $\lambda=2$ and $\lambda=0.5$, for $n=2$ and $n=3$. The coeffcients of both the short range approximation to $\eint$ and the long range approximation developed in \cite{spe-static} for these couplings are quoted in table \ref{tab1}. A more finely discretized dataset of these coefficients for $\lambda\in[0.1,2.5]$ can be found at \cite{data}.

\begin{table}[ht]
\begin{center}
\begin{tabular}{|c|c|cc|cc|}
 \hline
&  \textbf{$n=2$} & \multicolumn{2}{c|}{$n=3$} & \multicolumn{2}{c|}{\textbf{long-range}} \\
 \hline
$\lambda$& $c_2$ & $c_2$ & $c_3$  & $q$ & $m$ \\
 \hline
0.5& 0.0181513&	0.00399462&	0.00273742&	8.34655004&	13.83923926\\
1&	6.06E-09&	-1.93E-08&	6.12E-09&	10.72945106&	10.72878913\\
2&	-0.1070864&	-0.0238484&	-0.0348234&	15.24390759&	8.9584101\\
\hline
\end{tabular}
\end{center}
\caption{The coefficients $c_2^{n=2}$, $c_2^{n=3}$ and $c_2^{n=3}$ in the short range approximation to the vortex interaction energy for $n=2$ and $n=3$ vortices at couplings $\lambda=0.5$, $\lambda=1$ and $\lambda=2$. The data for $\lambda=1$ are included as a numerical check: all coefficients $c_k$ are known to vanish exactly in this case. The final two columns give the scalar monopole charge $q$ and the magnetic dipole moment $m$ of a single vortex, as used to compute the long range asymptotics of $\eint$ (see \cite{spe-static}). Again, the $\lambda=1$ data provide a numerical check, as it is known that $q=m$ exactly at critical coupling.}\label{tab1}
\end{table}

\begin{figure}
\begin{center}
\includegraphics[width=\linewidth]{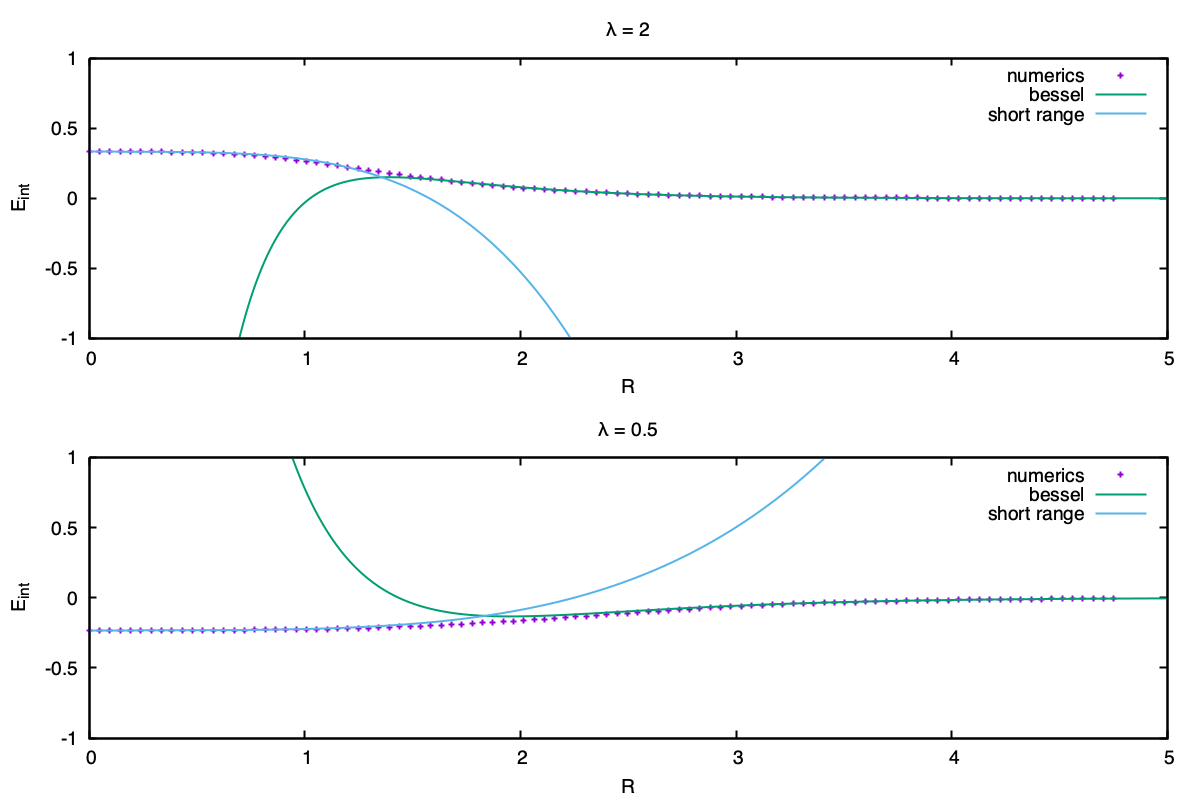}
\caption{Plot of numerical interaction energies (points) for two vortices of separation $2R$, compared with the approximation for the short range interaction in \eqref{eq:B2int} (blue) and the long range interaction given by the point source approximation in the linearized model (green).}
\label{fig:intB2}
\end{center}
\end{figure}

By translation and rotation invariance, $\eint$ for $n=2$ depends only on the distance between vortices, so it suffices to consider the one-parameter family of minimal energy configurations with vortices at $-R$ and $R$, for $R\geq 0$. As argued above, the correct coordinate on $\M_2^0$ is not the vortex separation $2R$, but rather the polynomial coeffcient $a_2=R^2$. Hence, our small $R$ approximation is
\beq
\eint(R) = E_2 - 2E_1 + \frac{c_2}{2}R^4.
\label{eq:B2int}
\eeq
Note that the two vortex interaction energy at short range is quartic, not quadratic, in $R$. This formula is compared with the numerically computed $\eint$ for $\lambda=2$ and $\lambda=0.5$ in figure \ref{fig:intB2}. The match is very close until it crosses with the long-range approximation
\cite{spe-static}
\beq
\eint(R)=\frac1{2\pi}\left[m(\lambda)^2K_0(2R)-q(\lambda)^2K_0(2\sqrt{\lambda}R)\right].
\eeq

\begin{figure}
\begin{center}
\includegraphics[width=\linewidth]{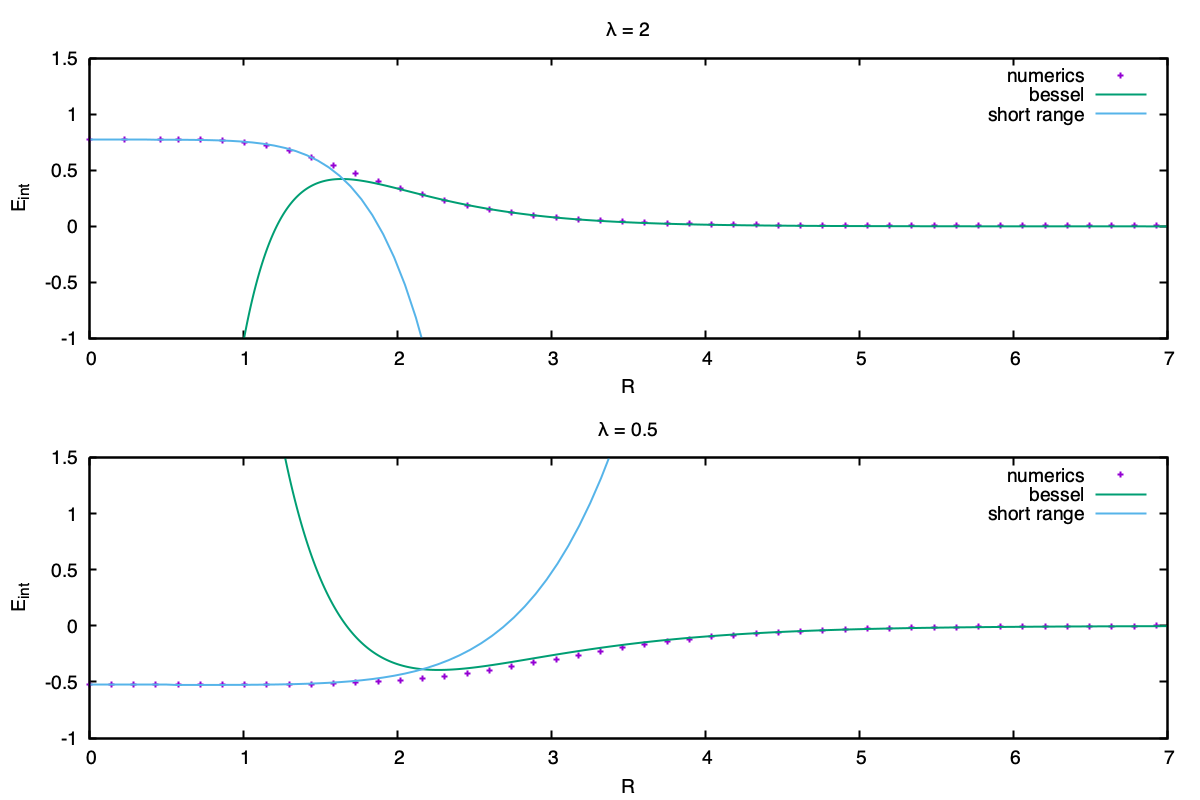}
\caption{Plot of numerical interaction energies (points) for three vortices in an equilateral triangle with distance $R$ from the origin, compared with the approximation for the short range interaction in \eqref{eq:short} (blue) and the long range interaction given by the point source approximation in the linearized model (green).}
\label{fig:intB3tri}
\end{center}
\end{figure}

\begin{figure}
\begin{center}
\includegraphics[width=\linewidth]{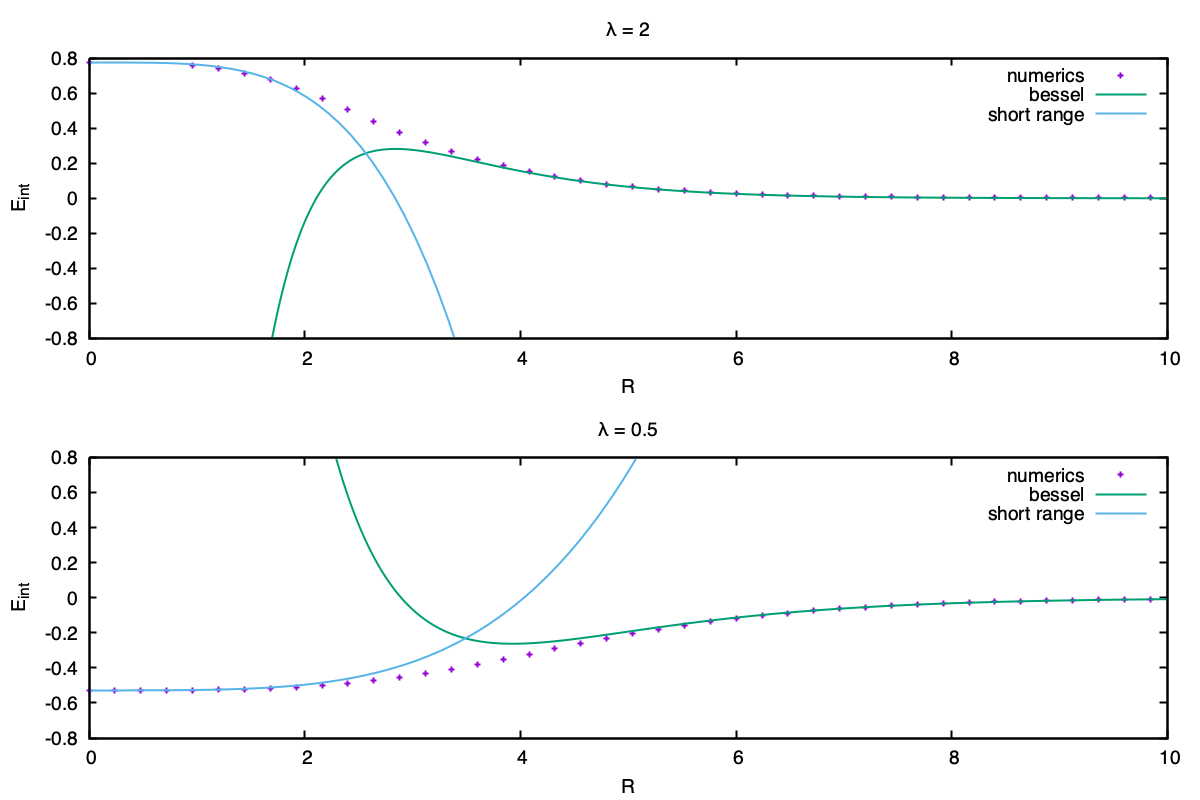}
\caption{Plot of numerical interaction energies (points) for three vortices in an equispaced line with distance to nearest neighbour $R$, compared with the approximation for the short range interaction in \eqref{eq:short} (blue) and the long range interaction given by the point source approximation in the linearized model (green).}
\label{fig:intB3line}
\end{center}
\end{figure}

The three vortex interaction energy is more complicated. By translation and rotation invariance it reduces to a function of $(|a_2|,a_3)\in[0,\infty)\times\C$. Rather than attempt to survey this entire three-dimensional space, we will compute the restriction of $\eint$ to 2 curves within it, namely, the curve 
\beq
p(z)=z^3-R^3
\eeq
consisting of vortices at the vertices $R$, $Re^{2\pi i/3}$, $Re^{-2\pi i/3}$ of an equilateral triangle and
\beq
p(z)=z^3-R^2z
\eeq
consisting of colinear vortices located at $-R$, $0$ and $R$. 
An immediate prediction of our short range approximation is that $\eint$ should be sextic in $R$ for the triangle curve and quartic for the line. A comparison with numerical data for $\lambda=2$ and $\lambda=0.5$ is given in figures \ref{fig:intB3tri} and \ref{fig:intB3line}.

One should note that the graph for the line of 3 vortices in the type II case ($\lambda=2$) only has field theory data for $R \geq 1$ due to a numerical artifact. For $R<1$ it becomes energetically favourable (on the lattice) for
the central zero to spread into a line from $-R$ to $R$ around which the phase of $\phi$ winds only once, and for $\phi$ to spawn two extra (winding $1$) zeroes towards the boundary of the computational domain.  We therefore removed these spurious data points. This pathology is absent in the triangular case because it is forbidden by symmetry, and absent always in the type I case for energetic reasons (it is never favourable to spawn extra well-separated zeroes).

\section{Concluding remarks}\news\label{sec:cr}

In this paper we have demonstrated that the spectral data of the Jacobi operator for the cocentred $n$-vortex can be used to infer the short range behaviour of the $n$ vortex interaction potential. This reduces a very challenging field theory problem (computing $\eint$ directly by constrained energy minimization) to a sequence of simple linear ODE problems. We have compared the resulting short range formulae to full field data for $n=2$ and $n=3$ in both the type I and type II regimes, finding good agreement up to vortex separations of around $3$. Remarkably, the range of validity of the short range approximation comes rather close overlapping the range of validity of the 
already established long range formulae \cite{spe-static}. It would be straightforward to splice these together, using spline interpolation, for example, to produce global explicit approximations for $\eint$, which may be of great practical utility in condensed matter physics. To facilitate this, we have computed the spectral coefficients $c_2$ (for $n=2$) and $c_2$, $c_3$ (for $n=3$), and the point vortex charges $q$, $m$, for a range of values of coupling $\lambda$. These data can be accessed at \cite{data}. 

The methods introduced here can be straightforwardly generalized to deal with multicomponent Ginzburg-Landau theory, in which one has several Higgs fields $\phi_1,\phi_2,\ldots,\phi_N$. A key new phenomenon in such models is type 1.5 superconductivity \cite{carbabspe}, in which vortices attract at long range but repel at short range. The method introduced here provides an easy and computationally efficient way of surveying the (very large) parameter space of these models for this phenomenon: one needs the longest length scale of the linearization of the model about the vacuum to be magnetic (or hybrid magnetic), and $c_2^{n=2}<0$. The first condition is checked by simple linear algebra, the second by solving the associated spectral problem. 

A second new phenomenon possible in multicomponent models is vortex core splitting \cite{garcarbabspe}: the model may admit potential or gradient coupling terms which favour the splitting apart of the zeros of the condensates, so that $\phi_1^{-1}(0)\neq \phi_2^{-1}(0)$. The minimal energy $n=1$ structures are then bound states of fractional flux vortices in the individual condensates, often termed ``skyrmions." Again, this phenomenon can be efficiently detected via the spectrum of $J$. The rotationally symmetric $1$-vortex is now a saddle point of $E$, so $J$ acquires a negative core-splitting mode. 

In the relativistic setting of the abelian Higgs model our results describe the interactions of static vortices. Our analysis used only those eigenmodes of $J$ which emerge from its kernel at critical coupling, the so-called splitting modes. There is another interesting eigenmode of $J$, for $n=1$, in the symmetry class $\CC_0^+$, called the shape mode \cite{alomigque}. This generates ``breathing" oscillations of the vortex \cite{alomig}. It has recently been shown that short range vortex interactions are modified considerably if these normal modes are excited \cite{krureewin,alamanmat}. Consequently, even at critical coupling (where $\eint$ vanishes indentically), vortices may attract when their individual shape modes are excited, leading to the formation of fluctuation-induced orbital bound states.

\subsection*{Acknowledgements}
This work was supported by the UK Engineering and Physical Sciences Research Council through grant EP/P024688/1. TW would like to thank the School of Mathematics at the University of Edinburgh for funding his postdoctoral position. The authors gratefully acknowledge several valuable conversations with Steffen Krusch.

%Bibliography generated by:
\bibliographystyle{stylefile}
\bibliography{bibliography}

%Bibliography generated by:
%\bibliographystyle{/apps/srv01/pmt/speight/prose/papers/BIBLIOGRAPHY/stylefile}
%\bibliography{/apps/srv01/pmt/speight/prose/papers/BIBLIOGRAPHY/bibliography}

%\input{mass_splitting.bbl}

\end{document}